\documentclass[a4paper,fleqn]{cas-dc}

\usepackage[authoryear,longnamesfirst]{natbib}

\def\tsc#1{\csdef{#1}{\textsc{\lowercase{#1}}\xspace}}
\tsc{WGM}
\tsc{QE}

\usepackage{xcolor,mathtools,booktabs,bigdelim,hyperref,ulem}

\newcommand{\norm}[1]{\Vert#1\Vert}
\DeclareMathOperator*{\mean}{mean}
\newcommand{\scalefactor}{}
\renewcommand{\cite}[1]{\citep{#1}}
\newcommand{\revision}[2]{#2}
\newcommand{\minor}[2]{#2}

\begin{document}
\let\WriteBookmarks\relax
\def\floatpagepagefraction{1}
\def\textpagefraction{.001}

\shorttitle{Physics-informed graph neural networks for flow field estimation in carotid arteries}

\shortauthors{Suk et al.}

\title [mode = title]{Physics-informed graph neural networks for flow field estimation in carotid arteries}



%

\author[1]{Julian Suk}[orcid=0000-0003-0729-047X]

\cormark[1]


\ead{j.m.suk@utwente.nl}



\affiliation[1]{organization={University of Twente},
            city={Enschede},
            country={The Netherlands}}

\author[1]{Dieuwertje Alblas}





\author[2,3]{Barbara A. Hutten}

\affiliation[2]{organization={Amsterdam University Medical Center},
            city={Amsterdam},
            country={The Netherlands}}

\affiliation[3]{organization={Research Institute Amsterdam Cardiovascular Sciences, Diabetes \& Metabolism},
	city={Amsterdam},
	country={The Netherlands}}

\author[2,3]{Albert Wiegman}
\author[1]{Christoph Brune}
\author[2]{Pim van Ooij}
\author[1]{Jelmer M. Wolterink}

\cortext[1]{Corresponding author}



\begin{abstract}
Hemodynamic quantities are valuable biomedical risk factors for cardiovascular pathology such as atherosclerosis. Non-invasive, in-vivo measurement of these quantities can only be performed using a select number of modalities that are not widely available, such as 4D flow magnetic resonance imaging (MRI). In this work, we create a surrogate model for hemodynamic flow field estimation, powered by machine learning. We train graph neural networks
that include priors about the underlying symmetries and physics, limiting the amount of data required for training. This allows us to train the model using moderately-sized, in-vivo 4D flow MRI datasets, instead of large in-silico datasets obtained by computational fluid dynamics (CFD), as is the current standard. We create an efficient, equivariant neural network by combining the popular PointNet++ architecture with group-steerable layers. To incorporate the physics-informed priors, we derive an efficient discretisation scheme for the involved differential operators. We perform extensive experiments in carotid arteries and show that our model can accurately estimate low-noise hemodynamic flow fields in the carotid artery. Moreover, we show how the learned relation between geometry and hemodynamic quantities transfers to 3D vascular models obtained using a different imaging modality than the training data. This shows that physics-informed graph neural networks can be trained using 4D flow MRI data to estimate blood flow in unseen carotid artery geometries.
\end{abstract}




\maketitle

\section{Introduction}
Cardiovascular disease is the leading cause of death worldwide. In many cases, it is characterised by an accumulation of plaque in the arterial wall, which results in narrowing of the blood vessel. In extreme cases, such stenosis causes a shortage of oxygenated blood supply to downstream organs which can lead to
myocardial infarction, peripherial artery disease or
ischemic stroke~\cite{SabaLoewe2023}. Ischemic stroke is often the result of atherosclerosis in the carotid arteries. Carotid artery geometry has been identified as an indicator for atherosclerosis~\cite{StreckerKrafft2020}  
and disturbed blood flow~\cite{LeeAntiga2008}. Arterial blood flow, together with derived quantities like wall shear stress and oscillatory shear index, has been shown to correlate with initiation and progression of atherosclerosis~\cite{CibisPotters2016}.  
Thus, insight into patient-specific blood flow is invaluable for diagnosis, treatment and prevention
of cardiovascular disease.

Information about patient-specific hemodynamics can be clinically obtained in-vivo, via invasive measurements such as pressure sensors in catheterisation or via non-invasive imaging such as Doppler ultrasound, particle image velocimetry~\cite{EngelhardVoorneveld2028} or 4D flow magnetic resonance imaging (MRI)~\cite{SabaLoewe2023}. In particular, 4D flow MRI allows for the quantification of directional blood flow and has emerged as a leading technique for in-vivo hemodynamics measurement~\cite{BissellRaimondi2023}.
However,
4D flow requires expensive, specialised software and expert knowledge, in particular for the setting of scan parameters.
More importantly, the number of patients for whom hemodynamics measurements could potentially provide additional diagnostic information commonly exceeds a hospital's capacity for performing 4D flow MRI, making the number of available scanners a bottleneck. Furthermore, 4D flow MRI is prone to measurement noise and inaccurate electrocardiogram (ECG) gating~\cite{BissellRaimondi2023}.

Alternatively, information about patient-specific hemodynamics can be obtained in-silico via computational fluid dynamics (CFD) based on anatomical computed tomography (CT)~\cite{LiOuyang2024} or MRI~\cite{TaylorPetersen2023}. CFD simulation based on 3D vascular models extracted from anatomical imaging is a powerful tool to estimate hemodynamic quantities in-silico. CFD allows for accurate, physics-conforming blood flow estimation.
Its downside are long runtimes and high computational demand, as well as sensitivity to modelling choices, such as discretisation and boundary conditions, making it difficult to compare results across practitioners. Indeed, it has been shown that there is high variability across CFD simulations
performed by different research groups~\cite{ValenSendstadBergersen2018}. This underlines the fact that while being a powerful model for blood flow estimation, CFD is not comparable to in-vivo measurement and must be validated via reference measurements, like 4D flow MRI. These limitations can be prohibitive for widespread use
in clinical practice.

\begin{figure*}
	\centering
	\includegraphics[width=\scalefactor\textwidth]{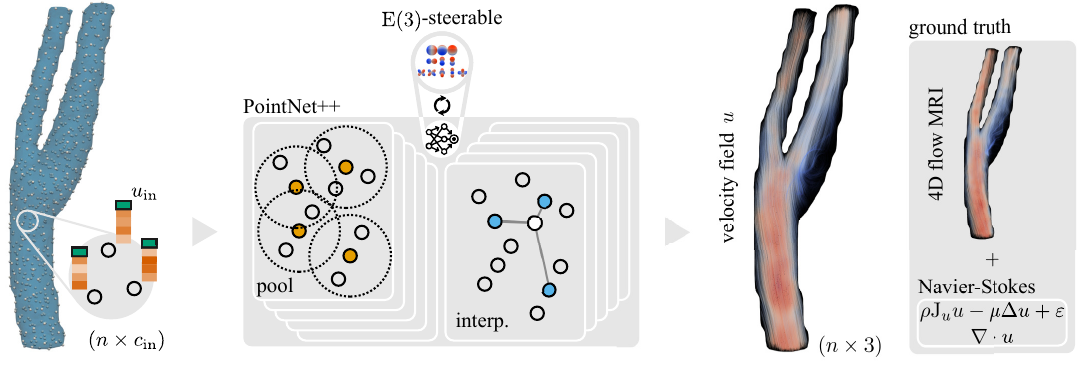}
	\caption{\textbf{Overview.} We represent the carotid artery by $n$ un-ordered points with $c_\text{in}$-dimensional input features, one of which is patient-specific inflow $u_\text{in}$ (shown in green). PointNet++ learns to map the input features to 3D velocity vectors using hierarchical down-sampling (pooling) and up-sampling (interpolation) layers. PointNet++ is comprised of learnable functions which we choose as vanilla multilayer perceptrons (MLP) or
	\revision{$\mathrm{E}(3)$-}{}steerable~\cite{BrandstetterHesselink2022} MLPs. We train PointNet++ with ground truth velocity fields obtained in-vivo via 4D flow MRI. Based on the Navier-Stokes equations for incompressible fluids that govern arterial blood flow, we regularise the training loss with residuals using discretised differential operators.
	}\label{fig:overview}
\end{figure*}

In this work, we aim to address the individual shortcomings of 4D flow MRI and CFD by creating an in-silico surrogate model for hemodynamic flow field estimation. Based on the observation that hemodynamics largely depend on artery shape and flow conditions, we propose a machine learning model that estimates patient-specific blood flow from personalised 3D vascular models. We demonstrate this approach in the carotid artery and train our machine learning model on 4D flow measurements in the carotid bifurcation of 234 subjects. We employ a powerful class of neural networks based on the PointNet++ architecture, which has been shown to excel at dealing with large point clouds~\cite{ZhangMao2023,MoralesFerezMill2021}. \revision{}{Notably, the model processes data in the form of point clouds rather than meshes to avoid overfitting to vertex connectivity and to overcome restrictions in receptive field.} Similarly as in CFD simulation~\cite{LopesPuga2020}, our machine-learning-based flow predictions are conditioned on patient-specific boundary conditions. Once trained, our model can quickly produce flow field estimates for vascular anatomy based on boundary conditions that can be cheaply obtained, e.g., via the widely available Doppler ultrasound. An overview of our method is provided in Fig.~\ref{fig:overview}.

Learning a relation between vascular geometry and hemodynamics from in-vivo data has the clear advantage that the estimated flow fields, unlike those obtained from CFD, do not depend on geometric and physical modelling choices. However, it poses the practical challenges that (1) training data is limited because 4D flow MRI is not performed on large cohorts of subjects and (2) measurement noise results in data pollution which can make training challenging. To address the former, we incorporate group equivariance by embedding steerable $\mathrm{E}(3)$-equivariant layers~\cite{BrandstetterHesselink2022}
in the PointNet++ architecture, which has been shown to facilitate extracting knowledge out of small hemodynamic datasets~\cite{SukBrune2023}. To address the latter, we include Navier-Stokes-based loss regularisation to mitigate the effect of un-physical noise in the training data. We perform experiments using conservation of mass as training loss regularisation alongside conservation of momentum\revision{ where we find it to be independently beneficial for accuracy}{}. We achieve this via discretisation of differential operators for which we derive a novel discretisation scheme using computationally efficient neighbourhood queries. The resulting model is fast, patient-specific and produces physics-conforming estimates.

The contributions of our work are as follows: (1) we propose a novel, powerful neural network for biomechanical surrogate modelling, (2) we propose an efficient way to include Navier-Stokes-based training regularisation and evaluation metrics and (3) we show how to train the machine learning (ML) model to perform volumetric velocity field estimation using in-vivo 4D flow MRI velocity fields and (5) we demonstrate how the learned relation between geometry and flow generalises to 3D shapes extracted from black-blood MRI, a different, much faster and widely-used MRI acquisition technique.

\section{Related works}\label{sec:relatedworks}
Recently, considerable attention has been placed on machine learning methods for hemodynamics estimation. These approaches can be divided into two categories: (1)
transductive instance optimisation with physics-informed regularisation (similar in scope to CFD) and (2)
inductive, generalising feed-forward methods that learn a relation between anatomy and hemodynamics from training data. Methods in the first category can incorporate personalised boundary conditions but lack the ability to generalise and require re-training for each case. In such, they can be useful in settings where the hemodynamic parameters are partially known~\cite{RaissiYazdani2020,FathiPerezRaya2020,KontogiannisJuniper2022}. In contrast, methods in the latter category learn to generalise to unseen data and aim to create fast, compute-efficient surrogate models which may replace compute-intensive CFD.

In this work we focus on the second category: learning a relation between subjects' vascular geometry and hemodynamics using neural networks. These neural networks operate on 3D representations of the vessels and estimate blood-flow-related quantities on the vessel wall or flow fields in the interior of the vessel. Besides efforts to estimate surface quantities like wall shear stress~\cite{SukHaan2024} and endothelial cell activation potential~\cite{MoralesFerezMill2021}, previous works have focussed on the estimation of volumetric (vector) fields. Among these, \citet{LiangMao2020} and \citet{WangWu2023} trained fully-connected neural networks operating on 3D point-cloud representations of the carotid artery and thoracic aorta, respectively, to estimate pressure and velocity fields.
\citet{MaulZinn2023} employed an octree-based neural network followed by trilinear interpolation to learn a continuous solution operator for the Navier-Stokes equations
in synthetic vascular trees. \citet{LiWang2021}  
used a PointNet-like~\cite{QiSu2016} architecture to estimate pressure and velocity fields in the coronary artery and synthetic cerebral aneurysm, respectively.
\citet{ZhangMao2023} composed PointNet++~\cite{QiYi2017} and a Navier-Stokes physics-informed neural network (PINN) which learned to estimate the velocity field in 3D models of the abdominal aorta.
In an earlier study~\cite{SukBrune2023}, we used a multiscale steerable $\mathrm{E}(3)$-equivariant graph neural network (SEGNN)~\cite{BrandstetterHesselink2022} to estimate velocity fields
in synthetic coronary arteries.

All aforementioned works are supervised learning methods, which are trained using reference labels in the form of hemodynamic ground truth acquired in-silico via CFD. While the use of CFD as ground truth facilitates acquiring large amounts of labelled data, this also implies that a properly trained model can at best mimic the results of the used CFD solver. As stated above, CFD solutions are sensitive to inaccuracies in the input geometry and faithful modelling of the boundary conditions and depend on the employed discretisation scheme.

\section{Materials and methods}
\subsection{4D flow MRI dataset}
We selectively included data of 234 subjects. All subjects participated in a long-term follow-up study evaluating cardiovascular risk in patients with familial hypercholesterolemia (FH) in whom statin treatment was initiated in childhood and in their unaffected siblings~\cite{LuirinkWiegman2019}.
All procedures were approved by the local institutional review board (METC) of the Amsterdam University Medical Center and were carried out according to the declaration of Helsinki. In all subjects, 4D flow MRI and 3D black blood MRI volumes were acquired.

4D flow scans were acquired using the following parameters: TR = 7.8 ms, TE = 4.6 ms, flip angle = 8°, VENC = 150 $\frac{\text{cm}}{\text{s}}$, $0.8 \times 0.8 \times 0.8$ $\text{mm}^3$ spatial and 80 ms temporal resolution. Retrospective ECG-triggering was used for cardiac synchronisation. Phase-offset corrections were performed during reconstruction. The 4D flow scan was accelerated using a k-t undersampling scheme of factor R = 8 and a k-t PCA reconstruction~\cite{PedersenKozerke2009}. Images were reconstructed using CRecon (Gyrotools, Zurich, Switzerland) with a regularisation factor of $r = 0.1$. The total scan time was 10 minutes. An nnU-Net model~\cite{IsenseeJaeger2020} was used to automatically segment a region-of-interest from an estimated 3 cm below to 2 cm above the carotid bifurcation. Triangular meshes for the carotid arteries were obtained from this segmentation and used to mask the velocity vector data. We chose the peak systolic time\revision{-}{ }frame as target in this study, which was defined as the time\revision{-}{ }frame with the highest spatially averaged velocity.

\revision{}{To facilitate visualisation of the flow fields, we created volumetric meshes from the surface meshes using TetGen~\cite{Si2015} and interpolated the velocity vector data at the mesh vertex positions with weights proportional to the vicinity to four closest velocity vector data points for each mesh vertex respectively. The mesh vertices acted as ground truth to train and evaluate our model.}

Black blood MRI scans were performed using the following parameters: TR = 10 ms, TE = 3.4 ms, flip angle = 6° and 0.5 mm $\times$ 0.5 mm $\times$ 0.5 mm spatial resolution. The k-space was under-sampled R = 5 times with a Poisson disk pattern and reconstructed using a compressed sensing algorithm for which the previously developed PROUD scanner software patch was used~\cite{PeperGottwald2020,GottwaldPeper2020}. Reconstruction was carried out in \textsc{MATLAB}, using MRecon (Gyrotools, Zurich, Switzerland) and the BART reconstruction toolbox. In all 3D black blood MRI volumes, the internal carotid artery (ICA), external carotid artery (ECA) and common carotid artery (CCA) were automatically segmented using a previously developed deep learning-based algorithm~\cite{AlblasBrune2021}. Meshes were subsequently cut at 3 cm below and 2 cm above the carotid bifurcation to match the region-of-interest obtained from 4D flow MRI.

The final dataset consisted of 128 left and 154 right carotid arteries, for which paired meshes were available from 4D flow MRI and black blood MRI.
For both 4D flow and black blood MRI, we obtained centerline points by binning and averaging geodesic segments of the artery surface, computed via the vector heat method~\cite{CraneWeischedel2017}. Connecting these centerline points by line elements allows us to query the distance to the centerline and thus vessel radius using trigonometry.

\subsection{Machine learning model}
In the following section we introduce our machine learning model for learning to predict velocity fields in 3D artery models. We propose a neural network that maps a collection of input features, defined on a set of points, which describe (local) artery geometry within the artery lumen, to a collection of output features representing 3D velocity vectors of blood flow.
Note that neither rotation nor translation of the input geometry should influence the relative structure of the predicted vector field. We address this by a steerable, \revision{$\mathrm{SE}(3)$}{$\mathrm{E}(3)$}-equivariant neural network that processes descriptive, geometric features independent of orientation in ambient space. Furthermore, we include the Navier-Stokes equations as loss regularisation.

\subsubsection{Input features}\label{sec:input}
\paragraph{Geometric descriptors}
We describe arterial geometry locally by point-wise feature vectors $f^p \in \mathbb{R}^c$ for each point $p \in \mathcal{P}_0$, where $c$ is the channel size and $\mathcal{P}_0$ a finite set of points within the artery lumen. We describe each point by its relative position (compare \cite{SukBrune2023}) to the artery inlet, lumen wall, outlets and centerline. We additionally encode relative position to the ICA outlet, ECA outlet and append a field of zeros (or ones) if the artery is a left (or right) carotid artery (respectively). We do so to account for possible physiological differences with respect to upstream or downstream vasculature and their effect on the flow field.  
All these features represent relative positions and are thus invariant under translation of the 3D geometry.

\paragraph{Inflow conditioning}
Furthermore, we include boundary condition features, as constant \revision{}{1D} scalar fields extended over all points\revision{, namely mean velocity and standard deviation over the artery inlet}{. In particular, we use the norm of the mean velocity vector over the artery inlet and the norm of the standard deviation vector over the artery inlet}. \revision{The latter}{This} is to simulate velocity measurements in an axial slice of the artery. Such measurements are commonly used in CFD to get subject-specific flow boundary conditions, often via ultrasound~\cite{LopesPuga2020}. Conditioning on different inflow velocities is important because boundary conditions vary between subjects and greatly influence blood flow.

\subsubsection{Architecture}\label{sec:gnn}

\begin{figure}
	\centering
	\includegraphics[width=\columnwidth]{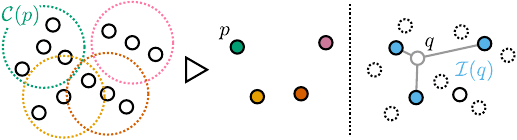}
	\caption{\textbf{Pooling and interpolation.} We use message passing layers to pool clusters $\mathcal{C}(p)$ of fine-scale to coarse-scale features in the contracting pathway (left) and interpolation to expand coarse-scale features back to original resolution (right). For simplicity, we visualise interpolation in 2D based on three closest points.}\label{fig:message_passing}
\end{figure}

We propose an encoder-decoder model (following PointNet++~\cite{QiYi2017}) with a contracting and expanding pathway. Given an un-ordered set $\mathcal{P}_0 = \{p^1, \dots, p^n\}$ of point coordinates $p = (p_1, p_2, p_3)^\mathsf{T} \in \mathbb{R}^3$ within a subject's artery lumen, we build a nested hierarchy of $h$ sub-sampled sets
\[
\mathcal{P}_0 \supset \dots \supset \mathcal{P}_h.
\]
Pair-wise, each coarse-scale point $p \in \mathcal{P}_{i + 1} \coloneqq \mathcal{P}_{i, \mathrm{coarse}}$ is assigned to a cluster of fine-scale points $\mathcal{C}(p) \subset \mathcal{P}_{i} \coloneqq \mathcal{P}_{i, \mathrm{fine}}$ (see Fig.~\ref{fig:message_passing}). We let the point features traverse the multiscale hierarchy ${\mathcal{P}_0 \supset \dots \supset \mathcal{P}_h}$ in a contracting and expanding pathway (see Fig.~\ref{fig:overview}) with skip connections $f^\mathrm{skip}$.

\paragraph{Contracting pathway}
We use message passing layers with learned functions $\phi$ for pooling from fine to coarse scales
\begin{align*}
	m^{p, q} &= \phi(f^q, q - p), &\text{(message from $q$ to $p$)}\\
	f^p &\leftarrow \bigoplus\limits_{q \in \mathcal{C}(p)} m^{p, q} &\text{(coarse point feature update)}
\end{align*}
where $\bigoplus$ is either the maximum or mean operator.

\paragraph{Expanding pathway}
We use interpolation layers with learned functions $\psi$ with which we unpool from the coarse- to the fine-scale points $q \in \mathcal{P}_{i, \mathrm{fine}}$
\[
f^q \leftarrow \psi\left(\frac{\sum\limits_{p \in \mathcal{I}(q)} \alpha_{p, q} f^p}{\sum\limits_{p \in \mathcal{I}(q)}\alpha_{p, q}}, f^\mathrm{skip}\right), \hspace{8pt} \alpha_{p, q} \coloneqq \frac{1}{\norm{p - q}_2^2 + \epsilon}
\]
where $\mathcal{I}(q) \subset \mathcal{P}_{i, \mathrm{coarse}}$ contains the four closest points to $q$ in $\mathcal{P}_{i, \mathrm{coarse}}$ (compare Fig.~\ref{fig:message_passing}) and $\epsilon$ is a small number.
The learned functions $\phi$ and $\psi$ are parametrised as multilayer perceptrons (MLP) composed of linear layers, activation functions and batch normalisation layers.

\subsubsection{Steerable equivariant PointNet++}
We introduce an equivariant extension of PointNet++ using steerable representations. Generally, a function is called equivariant to a symmetry group, if group actions applied to the inputs result in the equivalent transformation of the outputs. By choosing $\phi$, $\psi$ as steerable MLPs we, in turn, render our PointNet++ $\mathrm{O}(3)$-equivariant.  
Steerable MLPs that are equivariant under transformations of the orthogonal group $\mathrm{O}(3)$ (rotations and reflections) can be constructed by interleaving Clebsch-Gordan tensor products, gated activation functions and batch normalisation~\cite{BrandstetterHesselink2022}. They require expressing feature vectors throughout the network as steerable tensors (e.g. collections of scalars $s$ and 3D vectors $v$)
\[
f^p = (s_1, s_2, \dots, (v^1)^\mathsf{T}, (v^2)^\mathsf{T}, \dots)^\mathsf{T}
\]
whose rotation under $R \in \mathrm{O}(3)$ is well-defined:
\[
R \colon f^p \mapsto (s_1, s_2, \dots, (R v^1)^\mathsf{T}, (R v^2)^\mathsf{T}, \dots)^\mathsf{T}.
\]
We call this model steerable equivariant (SE-)PointNet++.


\subsubsection{Physics-based loss regularisation}\label{sec:pign}
Hemodynamics in the carotid arteries can be modelled by the Navier-Stokes equations. In particular, the divergence $\nabla \cdot u$ of the velocity field $u$ describes conservation of mass and is a recommended quality control in 4D flow MRI~\cite{BissellRaimondi2023}. Both conservation of mass and momentum can additionally function as training loss regularisation and accuracy metrics, promoting and measuring the conformity with the underlying physics.

In the context of numerical methods for partial differential equations (PDE) as well as machine learning, spatial discretisation of such differential operators is predominantly done via the finite element method~\cite{GaoZahr2022}  
or graph exterior calculus~\cite{ShuklaXu2022}. Both require a mesh of the \textit{volumetric}, spatial domain consisting of simplices (e.g. tetrahedra) which poses two problems for our application. Firstly, velocity field measurements from 4D flow MRI, which we use as training data, do not come with a volumetric mesh and algorithmic mesh creation is a non-trivial task.
Secondly, tetrahedral meshes typically feature a lot more tetrahedra than vertices. Finite-element-discretised operators are sparse matrices containing basis function coefficients defined on the mesh elements, which means their number of non-zero entries scales with the number of tetrahedra. This scaling is reflected in high memory footprint and compute overhead. To circumvent these challenges, we derive a novel, compute-efficient discretisation scheme, which uses neighbourhood queries.

\paragraph{Discretised Navier-Stokes equations}
In the Navier-Stokes equations for incompressible fluids with velocity field $u$, conservation of mass simplifies to the continuity equation
\[
\nabla \cdot u = \frac{\partial u_1}{\partial x_1} + \frac{\partial u_2}{\partial x_2} + \frac{\partial u_3}{\partial x_3} = 0.
\]
Given a discrete velocity field, we approximate the continuity equation using \eqref{eq:discretisation} which we will derive below. Let $u^p$ denote the velocity at $p \in \mathcal{P}_0$ and let $\mathcal{N}(p) \subset \mathcal{P}_0$ be a local neighbourhood around $p$. Then
\[
(\nabla \cdot u)^p \approx \frac{1}{|\mathcal{N}(p)|} \sum\limits_{q \in \mathcal{N}(p)} \frac{u^q - u^p}{\norm{q - p}_2} \cdot \frac{q - p}{\norm{q - p}_2}  
\]
where $\cdot$ denotes the dot product between vectors. Conservation of momentum (neglecting gravity) is described by the equation
\[
\rho (\frac{\partial u}{\partial t} + \mathrm{J}_u u) = -\nabla P + \mu \Delta u,
\]
where $\rho, \mu$ are density and dynamic viscosity, $P$ is the pressure field, $\mathrm{J}_u$ the Jacobian of $u$ and $\Delta$ the Laplacian operator. Analogously to the divergence operator, we approximate
\begin{align*}
	(\mathrm{J}_u)^p &\approx \frac{1}{|\mathcal{N}(p)|} \sum\limits_{q \in \mathcal{N}(p)} \frac{(u^q - u^p) (q - p)^\mathsf{T}}{\norm{q - p}_2^2} &\text{and}\\
	(\Delta u)^p &\approx \frac{1}{|\mathcal{N}(p)|} \sum\limits_{q \in \mathcal{N}(p)} \frac{((\mathrm{J}_u)^q - (\mathrm{J}_u)^p) (q - p)}{\norm{q - p}_2^2}. &
\end{align*}

\paragraph{Discretisation of derivatives}

\begin{figure}
	\centering
	\includegraphics[width=\columnwidth]{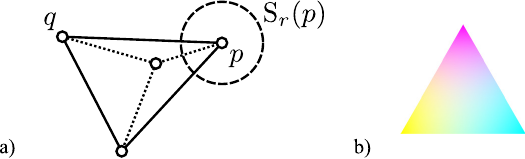}
	\caption{\textbf{Operator discretisation.} We visualise a) an element of a tetrahedral mesh used in our construction and b) colouring of a triangle using barycentric coordinates, as a visualisation of barycentric basis functions. Note that there is a linear gradient between just two colours along each of the three sides of the triangle. }\label{fig:operatordiscretisation}
\end{figure}

For the construction of our discretisation scheme, consider a tetrahedral mesh (which we will not actually have to create) endowed with a set of piecewise linear basis functions that allow barycentric interpolation of nodal degrees of freedom $u^p \in \mathbb{R}^3$
(see Fig.~\ref{fig:operatordiscretisation}~b)). Since the basis functions are piecewise linear, the resulting vector field $u \colon \mathbb{R}^3 \to \mathbb{R}^3$ is not differentiable at node locations. Consider a sphere $\mathrm{S}_r(p)$ of radius $r$ around $p$. We extend the derivative of $u$ to node locations $p$ by taking a surface integral over $\mathrm{S}_r(p)$ for some small $r$:
\begin{equation}\label{eq:surfaceintegral}
	\frac{\partial u_i}{\partial x_j}(p) \coloneqq \frac{1}{|\mathrm{S}_r(p)|} \int\limits_{\mathrm{S}_r(p)} \frac{\partial u_i}{\partial x_j}(s) \mathrm{d}s
\end{equation}
where $|\mathrm{S}_r(p)|$ denotes the surface area of the sphere. In the tetrahedral mesh, $p$ is connected to the vertex $q$ (see Fig.~\ref{fig:operatordiscretisation}~a)) by an edge $\xi = p + \lambda(\xi) (q - p)$, where $\lambda(x) = \frac{\norm{x - p}_2}{\norm{q - p}_2}$. Due to the barycentric basis functions, $u$ is linear on the edge:
\[
u_i(\xi) = u^p_i + \lambda(\xi) (u^q_i - u^p_i).
\]
By the chain rule, we have
\[
\frac{\partial u_i}{\partial x_j}(\xi) = \frac{\partial u_i}{\partial \lambda} \frac{\partial \lambda}{\partial x_j}(\xi) = \frac{u^q_i - u^p_i}{\norm{q - p}_2} \frac{\xi_j - p_j}{\norm{\xi - p}_2}.
\]
Now let $r < \lambda$. We can approximate the surface integral in \eqref{eq:surfaceintegral} by quadrature over the sphere where we let the grid points be induced by vertices from a neighbourhood $\mathcal{N}(p)$ around $p$:
\[
\int\limits_{\mathrm{S}_r(p)} \frac{\partial u_i}{\partial x_j}(s) \mathrm{d}s \approx \sum\limits_{q \in \mathcal{N}(p)} \frac{u^q_i - u^p_i}{\norm{q - p}_2} \frac{q_j - p_j}{\norm{q - p}_2}.  
\]
Substituting in \eqref{eq:surfaceintegral}, the derivatives become
\begin{equation}\label{eq:discretisation}
	\frac{\partial u_i}{\partial x_j}(p) \approx \frac{1}{|\mathcal{N}(p)|} \sum\limits_{q \in \mathcal{N}(p)} \frac{u^q_i - u^p_i}{\norm{q - p}_2} \frac{q_j - p_j}{\norm{q - p}_2}.
\end{equation}
For second-order derivatives we can derive a similar discretisation scheme. Assume, for a different set of basis functions, that $\frac{\partial u_i}{\partial x_j} \eqqcolon \partial_{x_j}u_i$ are piecewise linear on $\xi$:
\[
\partial_{x_j}u_i(\xi) = \partial_{x_j}u_i(p) + \lambda(\xi) (\partial_{x_j}u_i(q) - \partial_{x_j}u_i(p))
\]
Analogously to above, this yields the approximation
\[
\frac{\partial^2 u_i}{\partial x_j \partial x_k}(p) \approx \frac{1}{|\mathcal{N}(p)|} \sum\limits_{q \in \mathcal{N}(p)} \frac{(\partial_{x_j}u_i(q) - \partial_{x_j}u_i(p))}{\norm{q - p}_2} \frac{q_k - p_k}{\norm{q - p}_2}.
\]
Note that these discretisation schemes are mesh-free and can be computed at any point in space using simple neighbourhood queries. Furthermore, their computational complexity depends linearly on the number of vertices.

\paragraph{Overall training loss}
With the above, we can define
\begin{equation}\label{eq:continuity}
	L_\mathrm{continuity} \coloneqq \mean\limits_{p \in \mathcal{P}_0} |(\nabla \cdot u)^p|
\end{equation}
and
\begin{equation}\label{eq:momentum}
	L_\mathrm{momentum} \coloneqq \mean\limits_{p \in \mathcal{P}_0} \norm{\rho (\mathrm{J}_u)^p u^p - \mu (\Delta u)^p}_2  
\end{equation}
We combine these with $\mathrm{L}^1$ loss for the data term in the training loss. We balance the loss terms by multiplying a constant\revision{, such that loss term values roughly coincide at model initialisation}{based on observed training convergence}.

\subsection{Evaluation metrics}
To train and evaluate our neural network, we must compare velocity vector fields, i.e., point-wise model predictions $u^p$ and ground truth $\bar{u}^p$. We do so via the following metrics. Given an unordered set of points $\mathcal{P}_0$, we define approximation disparity
\[
\text{Approx. disp.}\colon \sqrt{ \sum_{p \in \mathcal{P}_0} \norm{\bar{u}^p - u^p}_2^2 / \sum_{p \in \mathcal{P}_0} \norm{\bar{u}^p}_2^2 }
\]
which measures the similarity between two vector fields. Furthermore, we use mean cosine similarity of two vector fields
\[
\text{Cos. similarity}\colon \mean_{p \in \mathcal{P}_0} \cos(\angle \bar{u}^p, u^p)
\]
which ranges between -1 (opposite)
and 1 (proportional) and measures directional agreement independent of magnitude.

\subsubsection{Regularised optimal transport distance}\label{sec:wasserstein}
Both approximation disparity and cosine similarity are useful metrics if point-wise comparison between velocity fields is possible. Beyond this case, we can only rely on comparison between local neighbourhoods of velocity vectors \revision{}{after spatial alignment (registration) of the point clouds}. A metric measuring the difference between local sets of vectors may be more robust to extreme outliers (as in noisy measurements) and can be used to compare discrete flow fields where the spatial positions do not coincide. To this end, we propose to use a regularised optimal transport distance, measuring the divergence between distributions of vectors. We compute the metric using an efficient implementation of the Sinkhorn divergence~\cite{FeydySejourne2019}  
for approximating optimal transport cost, which is effectively a relaxation of the \textit{Wasserstein distance}. \revision{}{Local neighbourhoods of points are constructed by first aligning both point clouds via Procrustes analysis and then including ground truth positions within a certain radius around the positions of model prediction.} Our metric takes into account the spatial position as well as direction and magnitude of two velocity vector fields.

\begin{figure*}
	\centering
	\includegraphics[width=\scalefactor\textwidth]{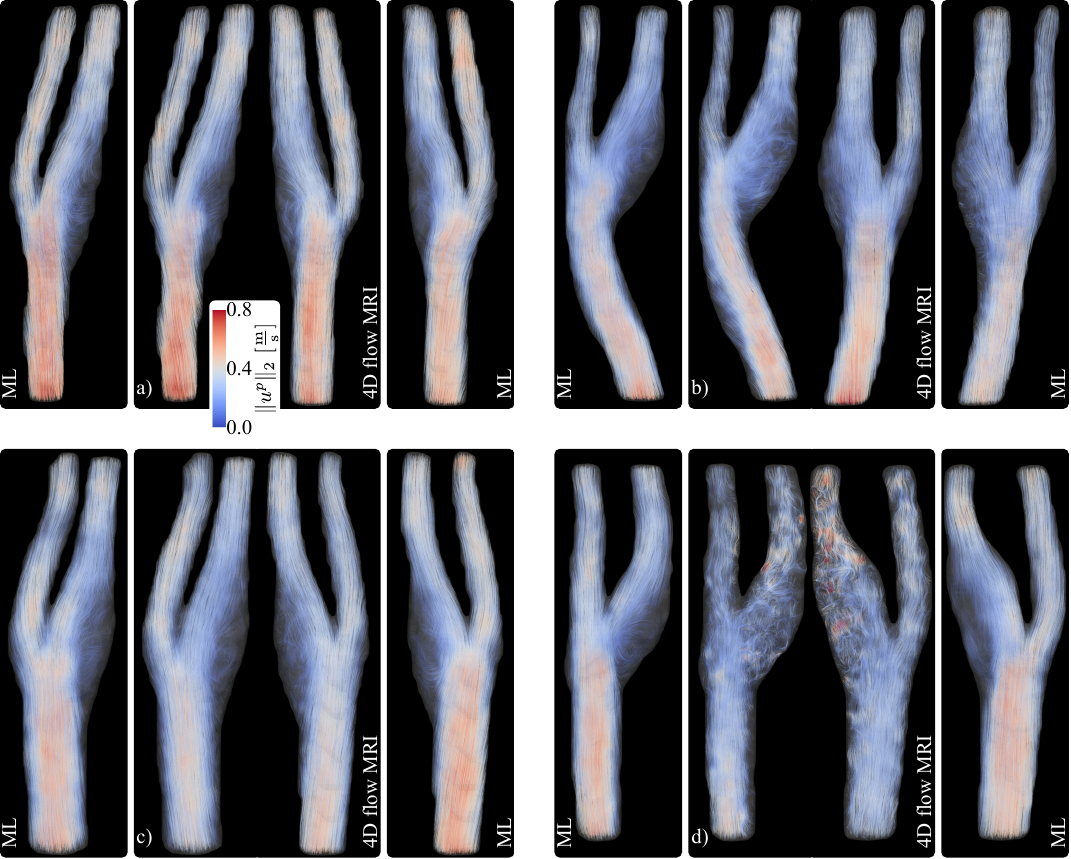}
	\caption{\textbf{Results} of neural network predictions (``ML") by PointNet++ compared with 4D flow MRI in the left and right carotid artery of subjects in the test split. We visualise the velocity field via 3D streamlines and render a projection image. Shown are examples of relatively a) good, b) average and c) poor performance. Additionally, we d) show an example where the ground truth is noisy, yet the model predicts a visually sound velocity field.}\label{fig:qualitative}
\end{figure*}

\begin{table*}
	\caption{\textbf{Quantitative} evaluation with respect to the 4D flow MRI measurements and flow physics (Navier-Stokes), evaluated via 10-fold cross-validation. We show mean $\pm$ standard deviation (across subjects) of the statistics. Since there is no point-to-point correspondence between black blood MRI and 4D flow MRI, Approx. disp. and Cos. similarity are omitted. We denote in \textbf{bold} superior performance where it significantly ($p < 0.05$) differs according to a one-way ANOVA test\revision{.}{: per metric, we picked the model with the lowest mean and tested significance compared to all other models. We consider all models whose metric does not differ significantly as equal.}}\label{tab:quantitative}
	\resizebox{\textwidth}{!}{
		\centering
		\renewcommand{\arraystretch}{1.3}
		\begin{tabular}{@{}llccccc@{}}
			\toprule
			Imaging modality & Neural network & Wasserst. dist. $\cdot \texttt{1e3}$ $\downarrow$ & Approx. disp. $\downarrow$ & Cos. similarity $\uparrow$ & $L_\mathrm{continuity}$ $\downarrow$ & $L_\mathrm{momentum} \cdot \texttt{1e-3}$ $\downarrow$ \\
			\midrule
			\multirow{7}{*}{4D flow MRI} & MLP & 17.6 $\pm$ 42.0 & 0.59 $\pm$ 0.16 & 0.72 $\pm$ 0.14 & 16.0 $\pm$ 1.6\hphantom{0} & 4.1 $\pm$ 0.5 \\
			& PointNet++$^*$ & 18.0 $\pm$ 41.2 & 0.61 $\pm$ 0.15 & 0.71 $\pm$ 0.14 & 16.3 $\pm$ 1.1\hphantom{0} & 4.8 $\pm$ 0.6 \\
			& PointNet++$^\dagger$ & 18.0 $\pm$ 42.4 & 0.58 $\pm$ 0.17 & \textbf{0.73} $\pm$ 0.14 & 16.6 $\pm$ 1.3\hphantom{0} & 5.2 $\pm$ 0.7 \\
			\cmidrule{2-7}
			& PointNet++ & 17.0 $\pm$ 41.7 & \textbf{0.56} $\pm$ 0.16 & \textbf{0.73} $\pm$ 0.14 & 12.5 $\pm$ 1.1\hphantom{0} & 3.7 $\pm$ 0.6 \\
			& SE-PointNet++ & 18.2 $\pm$ 43.9 & \textbf{0.55} $\pm$ 0.16 & \textbf{0.74} $\pm$ 0.14 & 9.8 $\pm$ 1.0 & 2.9 $\pm$ 0.5 \\
			\cmidrule{2-7}
			& PointNet++$^{\mathrm{PIGN}}$ & 17.6 $\pm$ 43.2 & \textbf{0.53} $\pm$ 0.16 & \textbf{0.74} $\pm$ 0.14 & 7.2 $\pm$ 0.7 & 2.5 $\pm$ 0.4 \\
			& SE-PointNet++$^{\mathrm{PIGN}}$ & 18.7 $\pm$ 44.5 & \textbf{0.54} $\pm$ 0.16 & \textbf{0.75} $\pm$ 0.14 & \textbf{5.8} $\pm$ 0.7 & \textbf{2.0} $\pm$ 0.4 \\
			\midrule
			\multirow{4}{*}{Black-blood MRI} & PointNet++ & 22.8 $\pm$ 48.3 & -- & -- & 16.1 $\pm$ 1.3\hphantom{0} & 4.6 $\pm$ 0.6 \\
			& SE-PointNet++ & 24.3 $\pm$ 50.5 & -- & -- & 11.6 $\pm$ 0.9\hphantom{0} & 3.1 $\pm$ 0.5 \\
			\cmidrule{2-7}
			& PointNet++$^{\mathrm{PIGN}}$ & 23.3 $\pm$ 50.0 & -- & -- & 10.4 $\pm$ 0.9 & 3.2 $\pm$ 0.5 \\
			& SE-PointNet++$^{\mathrm{PIGN}}$ & 24.5 $\pm$ 51.6 & -- & -- & \textbf{7.8} $\pm$ 0.6 & \textbf{2.3} $\pm$ 0.3 \\
			\bottomrule
			& \multicolumn{5}{l}{$^*$no geometric input features, only message passing, $^\dagger$no inflow conditioning}
		\end{tabular}
	}
\end{table*}

\section{Experiments and results}
All neural networks were implemented in Python using PyTorch~\cite{PaszkeGross2019} and PyTorch Geometric~\cite{FeyLenssen2019}, had approximately 1 million trainable parameters and were trained on an NVIDIA A100 (40 GB) GPU for 1000 epochs using Adam optimiser (learning rate $8 \cdot 10^{-4}$ with exponential decay of 0.9955) with gradient clipping.
Training PointNet++ took 11:13 s per epochs compared to 19:20 s for the equivariant SE-PointNet++. Runtime overhead of the physics-informed loss regularisation was negligible. The regularised Wasserstein distance computation (see Section~\ref{sec:wasserstein}) was implemented via GeomLoss~\cite{FeydySejourne2019}. We evaluated our neural networks via cross-validation. Our implementation is publicly available.\footnote{\href{https://github.com/sukjulian/physics-informed-gnn}{github.com/sukjulian/physics-informed-gnn}}

We partitioned the 282 subject-specific 4D flow MRI measurements \revision{ten}{10}-fold into 254 training and 28 evaluation samples, so that all subjects appeared in exactly one evaluation split (excluding two remainders). Left and right carotid artery of the same subject were considered separate samples, but we made sure that both were contained in the same fold. \revision{}{For training curves, see Appendix~\ref{app:convergence}.}

\subsection{Velocity field estimation in 4D flow MRI}\label{sec:results}

\subsubsection{PointNet++}
In Table~\ref{tab:quantitative} we compare test split predictions $u^p$ of PointNet++ to the corresponding 4D flow MRI measurements $\bar{u}^p$ via Wasserstein distance (see Section~\ref{sec:wasserstein}), approximation disparity
and cosine similarity
across all test-split subjects simultaneously.\footnote{\revision{}{For explanation of the high standard deviation in Wasserstein distance see Appendix~\ref{app:wasserstein}.}} Since some of the 4D flow MRI measurements were noisy (compare Fig.~\ref{fig:qualitative}~d)) we assess correspondence to the Navier-Stokes equations via continuity~\eqref{eq:continuity} and momentum residual~\eqref{eq:momentum}. PointNet++ achieves good correspondence to 4D flow MRI, indicated by cosine similarity close to one.
Fig.~\ref{fig:qualitative} visualises the reference velocity field computed via 4D flow MRI in the left and right carotid artery of four subjects from the combined cross-validation test splits. Comparison with the estimated velocity fields by PointNet++ shows good overall qualitative agreement to the reference. Even for the case where the 4D flow MRI velocity field is noisy (Fig.~\ref{fig:qualitative}~d)) PointNet++ produces a plausible estimate.

\subsubsection{SE-PointNet++}
\begin{figure}
	\centering
	\includegraphics[width=\columnwidth]{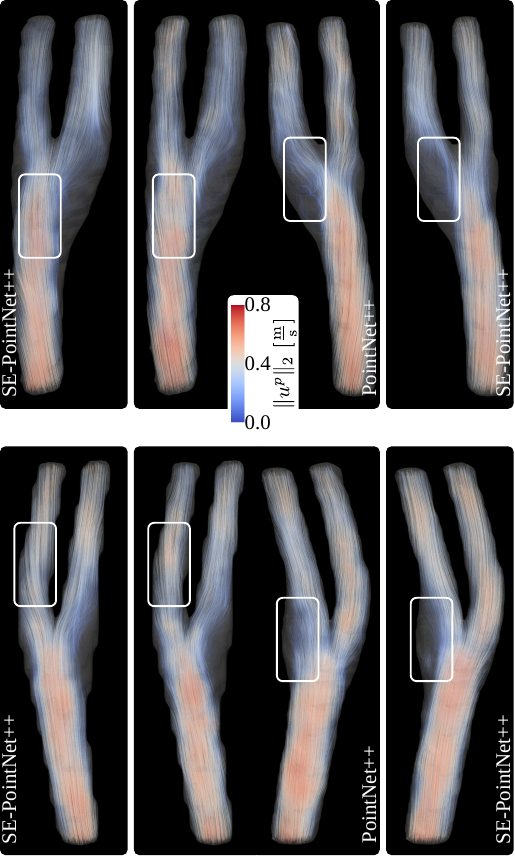}
	\caption{\textbf{Effect of group equivariance} of SE-PointNet++ compared to PointNet++. We show the left and right carotid artery of subjects in the test split and visualise the velocity field via streamlines.}\label{fig:equivariance}
\end{figure}

\revision{SE-PointNet++ obtained slightly higher approximation disparity than PointNet++, but we did not find the difference to be statistically significant ($p < 0.05$) in a one-way ANOVA test.}{SE-PointNet++ obtained the same approximation disparity as PointNet with no statistically significant ($p < 0.05$) difference in a one-way ANOVA test.}
SE-PointNet++ estimates had better conservation of mass and momentum than PointNet++ which is also supported by visual inspection.
Fig.~\ref{fig:equivariance}  shows a comparison between velocity field estimates by PointNet++ and by SE-PointNet++ and thus the influence of group equivariance. We found that SE-PointNet++ produced slightly smoother estimates while skipping over noise present in the PointNet++ estimate, e.g. in the ICA of the right carotid artery in Fig.~\ref{fig:equivariance}~a).

\subsubsection{Physics-informed extensions}
\begin{figure*}
	\centering
	\includegraphics[width=\textwidth]{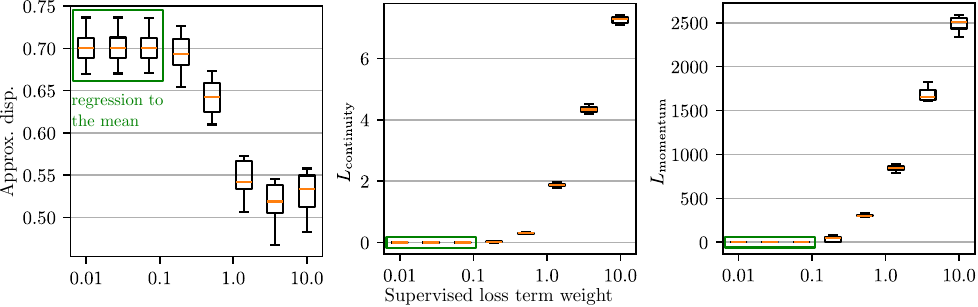}
	\caption{\revision{}{\textbf{Influence of supervised loss} in the presence of physics-informed loss regularisation. We trained PointNet++ for different supervised loss term weighting factors and computed Approx. disp. w.r.t. the 4D flow MRI data as well as $L_\mathrm{continuity}$ and $L_\mathrm{momentum}$ (corresponding to the physics-informed loss regularisation). In each case, we performed 10-fold cross-validation resulting in distributions which we visualise via boxplot.}}\label{fig:blend}
\end{figure*}

\begin{figure}
	\centering
	\includegraphics[width=\columnwidth]{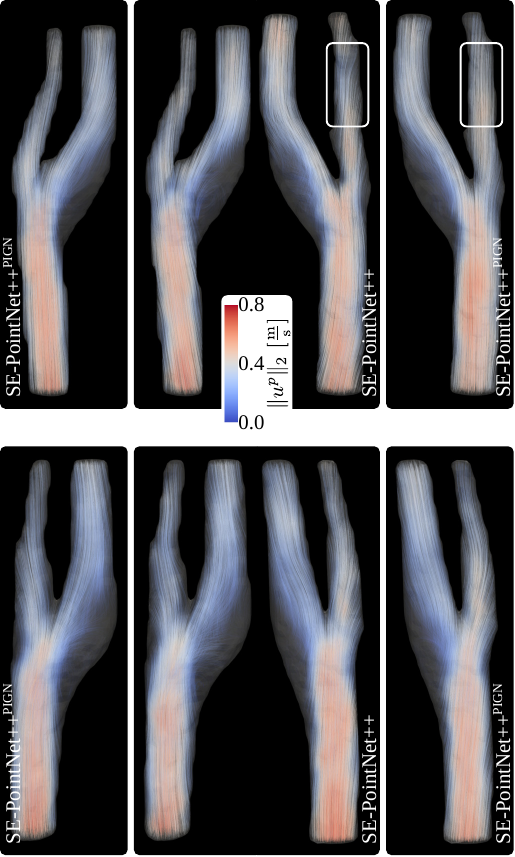}
	\caption{\textbf{Effect of physics-informed} training (``PIGN") for the example of SE-PointNet++. We show the left and right carotid artery of subjects in the test split and visualise the velocity field via streamlines.}\label{fig:pign}
\end{figure}

Using physics-informed loss regularisation (Section~\ref{sec:pign}), we found that continuity and momentum residuals decreased even further.\footnote{\revision{}{We weighted continuity and momentum terms by powers of ten (constant during training) to have two and four orders of magnitude lower values, respectively, than the data term. This was based on observed training convergence on the first out of 10 cross-validation splits.}} \revision{Additionally, we observed a slight, albeit statistically insignificant, decrease in approximation disparity for both neural networks.}{Decrease in approximation disparity was statistically insignificant for both neural networks.} \revision{We conclude that physics-based loss regularisation is independently beneficial for accuracy, i.e., does not compete with the training objective of fitting the data.}{} SE-PointNet++$^\text{PIGN}$ performs best w.r.t. correspondence to the continuity and momentum residual.
\revision{}{To investigate the interplay of the physics-informed loss regularisation and the supervised loss, we sampled eight weighting factors between 0.01 and 10 on a logarithmic scale and trained PointNet++ under differently weighted supervised loss term. In Fig.~\ref{fig:blend} we present approximation error (measuring the data fidelity) as well as the Navier-Stokes residuals as distributions over the cross-validation folds. We observed that a decrease in weighting factor for the supervised loss term correlated with an increase in approximation disparity (worse data fidelity) and a decrease in Navier-Stokes residuals (better adherence to the physics). The latter is due to the higher relative weighting of the physics-informed loss terms. Upon visual inspection, we found that when the physics-informed loss regularisation dominates (weighting factors $< 0.2$) the model regresses to the mean velocity vector everywhere in space. Note that a constant flow field minimises both $L_\mathrm{continuity}$ and $L_\mathrm{momentum}$.}
In Fig.~\ref{fig:pign} we compare velocity field estimates of SE-PointNet++ trained with and without physics-informed loss regularisation (SE-PointNet++$^\text{PIGN}$) based on the Navier-Stokes equations, as described in Section~\ref{sec:pign}. We found that this further smoothed the flow field, e.g. in the ECA of the right carotid artery in Fig.~\ref{fig:pign}~a).

\subsubsection{Ablation studies}
We created three ablated models to investigate the task-specific drivers of performance in PointNet++.
Firstly, we trained a simple MLP that maps the input features of each point to a velocity vector, without local or global interaction between points. \revision{}{In this ablation, all points are treated independently only based on their (input) features and the model does not have spatial context.} Secondly, we trained an ``empty" PointNet++ without geometric input features, i.e. point-wise description of relative geometry (see Section~\ref{sec:input}), which learned only from message passing between points (and boundary conditions). Thirdly, we trained a PointNet++ with geometric input features but without inflow conditioning (see Section~\ref{sec:input}).
Performance metrics are given in Table~\ref{tab:quantitative}. We found that all three variant achieved significantly worse approximation disparity than the baseline PointNet++. \revision{MLP achieved lower approximation disparity than the ``empty" PointNet++ pointing to the importance of the geometric input features.}{MLP and the ``empty" PointNet++ did not have statistically significant difference in approximation disparity, pointing to equal importance of geometric input features and spatial context.} Training time per epoch was 11:02 s for the MLP and thus runtime was comparable to PointNet++. Furthermore, we found that without inflow conditioning, PointNet++ significantly dropped in accuracy. This is expressed in higher approximation disparity but not lower cosine similarity, pointing towards magnitude rather than direction of predicted estimates.

%
%

\subsubsection{\revision{}{Subject-specific physics-informed graph neural network(s)}}
\revision{}{As additional reference for comparison besides 4D flow MRI, we trained -- for each subject -- a physics-informed graph neural network leveraging our physics-informed loss regularisation. We used PointNet++ for fast training and weighted the data term by 4, the continuity term by 0.01 and the momentum term by \texttt{1e-5} based on observed training convergence. Compared to the 4D flow MRI, the resulting model had mean ($\pm$ standard deviation) \text{Approx. disp.} of 0.29 ($\pm$ 0.14) across subjects. Physics-informed residuals $L_\mathrm{continuity}$ and $L_\mathrm{momentum} \cdot \texttt{1e-3}$ were 1.7 ($\pm$ 0.6) and 2.8 ($\pm$ 1.4), respectively. In Table~\ref{tab:pign} we present approximation disparity of all models (trained on 4D flow MRI) compared to the physics-informed graph neural network. We found the same relative accuracy as in Table~\ref{tab:quantitative} with the only difference that PointNet++ was slightly less accurate than its physics-informed counterpart here.}

\subsection{Generalisation to black-blood MRI}\label{sec:black_blood_mri}
\begin{figure*}
	\centering
	\includegraphics[width=\scalefactor\textwidth]{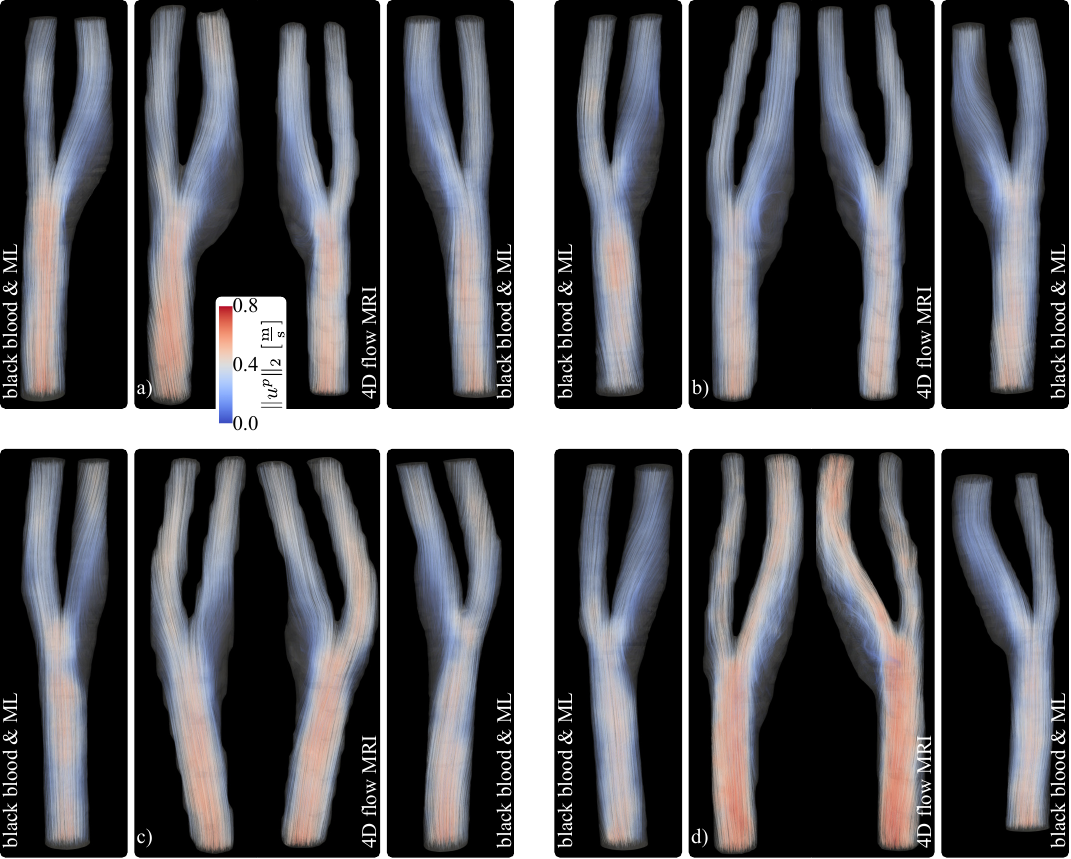}
	\caption{\textbf{Generalisation to black-blood MRI} in the left and right carotid bifurcation of subjects from the test split. We compare 4D flow MRI with application of a pre-trained machine learning model on geometries obtained via black blood MRI (denoted ``black blood \& ML"). Note that black blood MRI itself does not provide measurements of the flow field.}\label{fig:black_blood_mri}
\end{figure*}

We investigate our neural networks' capabilities to generalise between vascular models obtained using different imaging protocols, by training the neural networks on the 4D flow MRI dataset and letting them estimate velocity fields in 3D models of the \textit{same artery} but obtained via black-blood MRI. We pass the inflow boundary conditions extracted from the inlet region of the 4D flow MRI measurements (see Section~\ref{sec:input}) to the machine learning model, simulating in-vivo measurements via e.g. ultrasound. Note that black blood MRI in itself does not allow for velocity measurements, since it is not a temporal imaging method and does not measure blood flow or contains flow-encoding sequence parts.
Since black-blood MRI by itself does not allow for velocity measurements, we quantify our neural networks' accuracy by Wasserstein distance (see Section~\ref{sec:wasserstein}) to the patient-specific 4D flow measurements as well as continuity and momentum residual (see Section~\ref{sec:pign}). Table~\ref{tab:quantitative} lists the results. \revision{PointNet++ achieves the lowest mean Wasserstein distance, but the differences are not significant.}{Differences in Wasserstein distance between the models were not significant.} As above, SE-PointNet++$^\text{PIGN}$ achieves the lowest continuity and momentum residuals. These are slightly higher than for the 4D flow MRI data. Nevertheless, $L_\mathrm{continuity}$ is substantially lower in flow field estimates produced by SE-PointNet++$^\text{PIGN}$ than in those of PointNet++, SE-PointNet++ and PointNet++$^\text{PIGN}$.
Fig.~\ref{fig:black_blood_mri} shows a comparison between ground truth 4D flow MRI and machine learning estimates (denoted ``black blood \& ML") in 3D carotid artery models of subjects from the test split obtained via black blood MRI. We use SE-PointNet++ for this, since it, unlike PointNet++, does not require alignment of the black blood MRI geometry with the 4D flow MRI due to its roto-translation equivariance. This is to simulate a realistic scenario where 4D flow MRI would not be available. We find good agreement of the machine learning model with the ground truth, e.g. in the left carotid artery in Fig.~\ref{fig:black_blood_mri}~a) and the right carotid artery in Fig.~\ref{fig:black_blood_mri}~b). For several cases the machine learning model underestimates the magnitude of the flow (Fig.~\ref{fig:black_blood_mri}~d)). Nevertheless, we find overall good qualitative agreement in the estimated and measured flow fields.

\begin{table}
	\caption{\revision{}{\textbf{Comparison} to subject-specific physics-informed graph neural network via 10-fold cross-validation. We show mean $\pm$ standard deviation (across subjects). We denote in \textbf{bold} superior performance where it significantly ($p < 0.05$) differs according to a one-way ANOVA test: per metric, we picked the model with the lowest mean and tested significance compared to all other models. We consider all models whose metric does not differ significantly as equal.}}\label{tab:pign}
	\centering
	\renewcommand{\arraystretch}{1.3}
	\begin{tabular}{@{}lc@{}}
		\toprule
		Neural network & Approx. disp. $\downarrow$ \\
		\midrule
		MLP & 0.54 $\pm$ 0.19 \\
		\midrule
		PointNet++ & 0.50 $\pm$ 0.18 \\
		SE-PointNet++ & \textbf{0.49} $\pm$ 0.18 \\
		\midrule
		PointNet++$^\text{PIGN}$ & \textbf{0.46} $\pm$ 0.18 \\
		SE-PointNet++$^\text{PIGN}$ & \textbf{0.47} $\pm$ 0.19 \\
		\bottomrule
	\end{tabular}
\end{table}

\section{Discussion and conclusion}
In this work, we present a novel deep learning approach for modelling hemodynamics by generating velocity fields for unseen subject-specific carotid artery geometries. We build upon the PointNet++~\cite{QiYi2017} architecture and infuse it with steerable equivariant~\cite{BrandstetterHesselink2022}
MLPs.
We show how this symmetry consideration implicitly leads to stronger adherence to the underlying flow physics. Our neural network architecture allows for global context aggregation via cascading of pooling and interpolation layers and bypasses the need for message passing for each point in the original resolution, which would be computationally prohibitive. Thus, we are able to effectively learn features mapped to the original point cloud at low computational cost. We complement our deep learning approach by including domain knowledge, specifically about the differential structure of the velocity field, by regularising the training loss with discretised residuals based on the Navier-Stokes equations.
To this end, we derive a computationally efficient, mesh-free discretisation scheme whose complexity is linear in the number of points. Ultimately, we demonstrate
how our method can learn to estimate velocity fields in unseen subjects after being trained exclusively on 4D flow MRI data.

The approach we propose addresses some important limitations of the state-of-the-art methods in machine-learning-based flow field estimation (Section~\ref{sec:relatedworks}). Compared to single-case instance optimisation methods for fluid flow estimation~\cite{RaissiYazdani2020,FathiPerezRaya2020,KontogiannisJuniper2022}, our approach is able to generalise beyond the training data while still respecting the governing PDEs. Compared to previous works on multiple-case generalising feed-forward methods~\cite{LiangMao2020,WangWu2023,MaulZinn2023,LiWang2021,ZhangMao2023,SukBrune2023},  
our model gains independence of the variability of the CFD-based training data, because we train it on in-vivo flow field measurements from 4D flow MRI.

Although we used 4D flow MRI training data to train and validate our method, we also demonstrate
that we can train machine learning models on 4D flow MRI velocity fields and transfer the learned relations to vascular models obtained from anatomical black blood MRI images. This could be valuable in practice, as anatomical imaging modalities might be more widely available than 4D flow MRI and 3D vascular models can be obtained from these modalities automatically, e.g., in black blood MRI~\cite{AlblasBrune2021}, CT~\cite{LiOuyang2024} and 3D ultrasound. We envision this approach to be used in clinical practice for fast estimation of patient-specific blood flow, but backed up by 4D flow MRI for cases in which complications are suspected.
This would still require personalised boundary conditions, which could be non-invasively obtained using, e.g., Doppler ultrasound.\footnote{\revision{}{Doppler ultrasound comes with its own set of drawbacks such as \minor{}{its restriction to 1D and its} dependency on the measurement angle and user.}} Note however, that even though these geometries are of the same patient, different imaging modalities produce slightly different geometries which is a source of error in our method.

In 4D flow MRI, velocity cannot be accurately computed very close to the artery wall which results in high levels of noise.
In our experiments, even though the training data was noisy in some cases, our neural networks did not reproduce the noise but implicitly smoothed over it in their estimations. We attribute this to the fact that the noise cannot be explained by the input anatomy in a meaningful way and our neural networks treat it as constant error term. \revision{This has potential implications for the estimation of wall shear stress which requires noise-free velocity estimates close to the artery wall.}{It would be interesting to investigate to what extent the estimated velocity field could be smoothed using a similar strategy as our physics-informed loss regularisation (Section~\ref{sec:pign}) and if this could enable extraction of wall shear stress which requires noise-free velocity estimates close to the artery wall.}

\revision{}{We found that our $\mathrm{E(3)}$-equivariant neural network \textit{a priori} produced estimates with significantly lower Navier-Stokes residuals compared to the vanilla baseline. This can be partly explained by the inherent symmetries present in fluid flow. However, this phenomenon should be systematically studied and is an interesting avenue for future work.}

Since our method is a data-driven approach, the quality of the results largely depends on the
extent to which the training data represents the data distribution during inference. \revision{}{For instance, a model trained on arteries of a certain topology (number of outles, bifurcations, etc.) is unlikely to generalise to substantially different topologies. This might also apply to scans with substantially different field of view (up- or downstream) dependent on the specific representation the model has learned (e.g. dependence on absolute coordinates).}
In contrast to comparable studies using synthetic flow data generated by CFD~\cite{LiangMao2020,WangWu2023,MaulZinn2023,LiWang2021,ZhangMao2023,SukBrune2023}, we have access to a small dataset in this work. This poses no major challenge because the carotid bifurcation across different subjects in our data \revision{were}{are} of a relatively similar shape. However, it would be interesting to see what happens for larger variations in the data\revision{.}{, both geometric and regarding flow data. This would especially be relevant for large datasets that are collaboratively sourced from multiple institutes: 4D flow MRI can vary significantly across methodologies and hardware. In such a scenario, velocity fields should be canonicalised for optimal model performance. In principle, we could also including well-tuned CFD simulations in our training data to complement the 4D flow MRI, as the former can also be represented as point clouds of velocity vectors. Such a hybrid-data approach is an interesting avenue for future research.} \minor{}{As data-drive approach, our method inherits the dependency on acquisition settings and image quality from the 4D flow MRI on which it is trained. Just like 4D flow MRI, comparability of the generated velocity fields with those acquired under different conditions by different clinicians is critical for potential clinical translation. Testing this mechanism with fixed patients would be crucial insight into the robustness of our method and is thus important future research.} Besides stenotic cardiovascular disease (CVD), aneurysmatic CVD, in the form of intracranial and abdominal aneurysms, is also a dangerous disease which depends on blood flow. Thus, aneurysmatic CVD also provides potential applications of our methods in determination of hemodynamic quantities in these kinds of diseases. Even though we focussed on the carotid arteries in this work, we expect our approach to apply to other arteries as well, as long as there is a sufficiently representative dataset relative to the artery complexity (e.g. stenoses, aneurysms or increased tortuosity). What is more, the quantitative evaluation in our experiments is held back by the measurement noise present in some of the test cases since it obfuscates the expressiveness of the evaluation metrics. We address this by performing statistical significance tests to discern differences in the numerical results.  
Lastly, our analysis in this work is limited to \revision{time-averaged}{single time frame} instead of pulsatile hemodynamics estimation, even though 4D flow MRI does enable transient quantification of blood flow. \minor{}{We recognise that single time frame analysis might obfuscate turbulence effects.} \revision{}{Especially in the context of stenotic CVD, turbulence plays a prominent role, modelling which is usually performed in transient flow. Incorporating an explicit turbulence model in our physics-informed loss regularisation is an interesting avenue for future research.}

In conclusion, physics-informed graph neural networks can be trained using 4D flow MRI data to cheaply estimate blood flow in new and unseen carotid artery geometries.

\section*{Acknowledgement}
We would like to thank José Iglesias Martínez for his help in the form of discussions about optimal transport and Sinkhorn divergence.
This work is funded in part by the 4TU Precision Medicine programme supported by High Tech for a Sustainable Future, a framework commissioned by the four Universities of Technology of the Netherlands. Jelmer M. Wolterink was supported by the NWO domain Applied and Engineering Sciences Veni grant (18192).

\appendix

\section*{Appendix}

\section{\revision{}{Standard deviation in Wasserstein distance}}\label{app:wasserstein}
\begin{figure}[h]
	\centering
	\includegraphics[width=.333\textwidth]{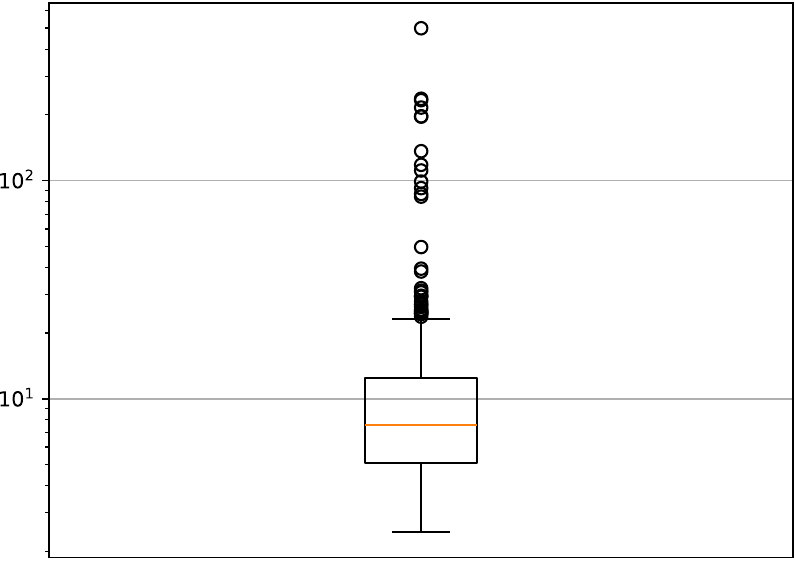}
	\caption{\revision{}{\textbf{Distribution} of Wasserstein distance for MLP compared to 4D flow MRI. The orange line denotes the median (Wasserst. dist. $\cdot \texttt{1e3}$ = 7.6). Outliers correspond to 4D flow MRI data with very low signal-to-noise ratio (compare Fig.~\ref{fig:qualitative}~d)). We did not exclude these outliers to simulate a realistic scenario.}}\label{fig:wasserstein}
\end{figure}

\revision{}{See Fig.~\ref{fig:wasserstein}.}

\section{\revision{}{Training convergence}}\label{app:convergence}
\begin{figure*}[h]
	\centering
	\includegraphics[width=\textwidth]{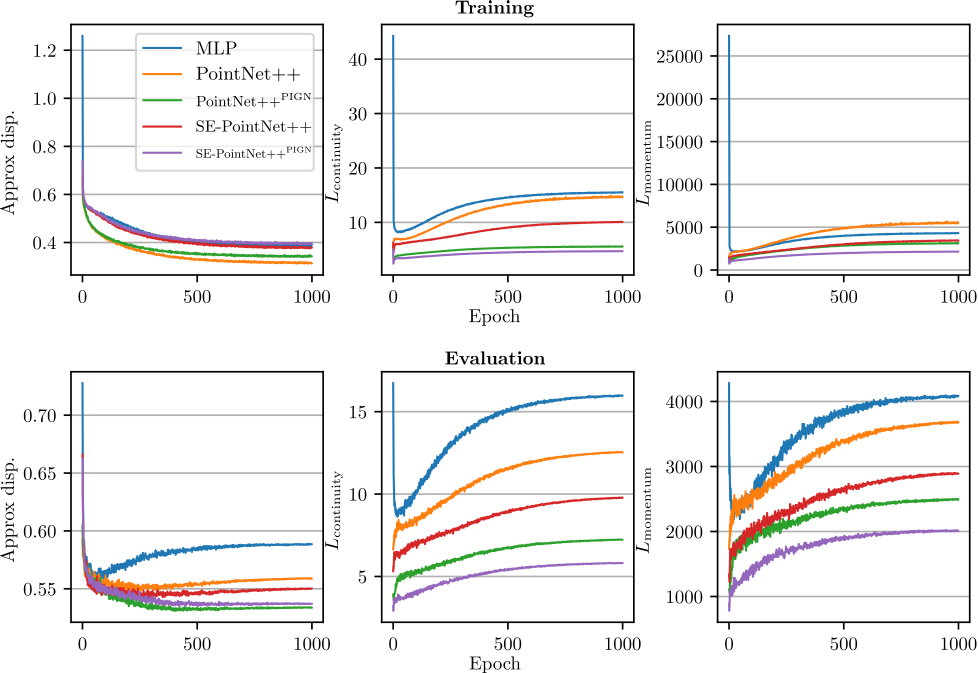}
	\caption{\revision{}{\textbf{Training curves} and evaluation metrics progression during training. We show mean Approx. disp., $L_\mathrm{continuity}$ and $L_\mathrm{continuity}$ over the cross-validation folds. Note that we did not have held-out validation splits for this experiment.}}\label{fig:convergence}
\end{figure*}

\revision{}{See Fig.~\ref{fig:convergence}.}

\bibliographystyle{cas-model2-names}

\bibliography{cas-refs}



\end{document}